\documentclass[twocolumn]{aastex631}

\received{December 28, 2021}
\accepted{January 18, 2022}
\acceptjournal{Astronomical Journal}

\shorttitle{WASP-77 A b Formed Beyond Its Parent Protoplanetary Disk's
H$_{2}$O Ice Line}
\shortauthors{Reggiani et al.}

\begin{document}

\title{Evidence that the Hot Jupiter WASP-77 A b Formed Beyond Its Parent
Protoplanetary Disk's H$_{2}$O Ice Line}

\correspondingauthor{Henrique Reggiani}
\email{hreggiani@carnegiescience.edu}

\author[0000-0001-6533-6179]{Henrique Reggiani}
\altaffiliation{Carnegie Fellow}
\affiliation{The Observatories of the Carnegie Institution
for Science, 813 Santa Barbara St, Pasadena, CA 91101, USA}

\author[0000-0001-5761-6779]{Kevin C. Schlaufman}
\affiliation{William H.\ Miller III Department of Physics and Astronomy, Johns Hopkins
University, 3400 N Charles St, Baltimore, MD 21218, USA}

\author[0000-0002-7718-7884]{Brian F. Healy} 
\affiliation{William H.\ Miller III Department of Physics and Astronomy, Johns Hopkins 
University, 3400 N Charles St, Baltimore, MD 21218, USA}

\author[0000-0003-3667-8633]{Joshua D. Lothringer}
\affiliation{Department of Physics, Utah Valley University, College of
Science - MS 179, 800 W University Pkwy, Orem, UT 84058, USA}

\author[0000-0001-6050-7645]{David K. Sing}
\affiliation{William H.\ Miller III Department of Physics and Astronomy, Johns Hopkins
University, 3400 N Charles St, Baltimore, MD 21218, USA}
\affiliation{Department of Earth and Planetary Sciences, Johns Hopkins
University, 3400 N Charles St, Baltimore, MD 21218, USA}

\begin{abstract}

\noindent
Idealized protoplanetary disk and giant planet formation models have been
interpreted to suggest that a giant planet's atmospheric abundances can
be used to infer its formation location in its parent protoplanetary
disk.  It has recently been reported that the hot Jupiter WASP-77 A
b has sub-solar atmospheric carbon and oxygen abundances with a solar
C/O abundance ratio.  Assuming solar carbon and oxygen abundances for
its host star WASP-77 A, WASP-77 A b's atmospheric carbon and oxygen
abundances possibly indicate that it accreted its envelope interior to
its parent protoplanetary disk's H$_{2}$O ice line from carbon-depleted
gas with little subsequent planetesimal accretion or core erosion.
We comprehensively model WASP-77 A and use our results to better
characterize WASP-77 A b.  We show that the photospheric abundances
of carbon and oxygen in WASP-77 A are super-solar with a sub-solar C/O
abundance ratio, implying that WASP-77 A b's atmosphere has significantly
sub-stellar carbon and oxygen abundances with a super-stellar C/O ratio.
Our result possibly indicates that WASP-77 A b's envelope was accreted
by the planet beyond its parent protoplanetary disk's H$_{2}$O ice line.
While numerous theoretical complications to these idealized models have
now been identified, the possibility of non-solar protoplanetary disk
abundance ratios confound even the most sophisticated protoplanetary
disk and giant planet formation models.  We therefore argue that giant
planet atmospheric abundance ratios can only be meaningfully interpreted
relative to the possibly non-solar mean compositions of their parent
protoplanetary disks as recorded in the photospheric abundances of their
solar-type dwarf host stars.

\end{abstract}


\section{Introduction}\label{intro}

It has been proposed that a giant planet's formation location in its
parent protoplanetary disk can be discerned by studying the abundances
of the elements in the planet's atmosphere.  While acknowledging the
importance of planetesimal accretion after runaway envelope accretion
in determining giant planet envelope abundances, based on static disk
models \citet{oberg2011} suggested that stellar C/O abundance ratios
should be realized for giant planets forming inside both their parent
protoplanetary disks' H$_{2}$O ice and carbon-grain sublimation lines
where all volatiles are in the gas phase.  Sub-stellar or stellar
C/O abundance ratios combined with super-stellar carbon abundances
were argued to result from the accretion of large amounts of icy
planetesimals after envelope accretion.  Super-stellar C/O abundance
ratios and carbon abundances could be attributed either to formation
close to the CO$_{2}$ or CO ice lines or by the accretion of carbon-rich
grains in the narrow range inside the H$_{2}$O ice line but outside the
carbon-grain sublimation line.  Super-stellar C/O abundance ratios and
sub-stellar carbon and oxygen abundances were put forward as a unique
signature of formation beyond the H$_{2}$O ice line.

This straightforward scenario outlined for the case of static disks
is significantly complicated in more realistic models that include
disk chemical \& structural evolution, the radial migration of solids,
detailed models for planetesimal accretion, and/or post envelope-accretion
migration.  The locations of the H$_{2}$O, CO$_{2}$, and CO ice lines
depend on stellar luminosity and move inward with time as a viscously
evolving disk's mass accretion rate declines \citep[e.g.,][]{harsono2015}.
The inclusion of radially migrating solids in viscously evolving disks
can also cause the H$_{2}$O ice line to move in by a factor of two
\citep{piso2015}.  In this case, the C/O abundance ratio in disk gas is
most likely to be super-stellar between the CO$_{2}$ and CO ice lines
\citep{piso2015}.  Gas drag, sublimation, vapor diffusion, condensation,
and coagulation on major carbon- and oxygen-bearing volatiles indicate
that giant planets that accreted their envelopes inside the H$_{2}$O ice
line should have super-stellar C/O \citep{ali-dib2014,ali-dib2017}.
In the absence of core erosion, pebble accretion is thought to
produce sub-stellar carbon and oxygen abundances but super-stellar C/O
abundance ratios \citep[e.g.,][]{madhusudhan2017,booth2017}.  The point
in time during a disk's evolution at which a giant planet forms also
affects the C/O abundance ratio of its envelope \citep{cridland2016}.
Chemical evolution in disks can mask the influence of formation location
on envelope composition though \citep{eistrup2018}, with super-stellar C/O
only realized between CO$_{2}$ and CH$_{4}$ ice lines and greater giant
planet envelope abundance dispersions expected in more metal-rich disks
\citep{cridland2020,notsu2020}.  High-eccentricity migration after disk
dissipation can lead to sub-stellar carbon and oxygen abundances with
stellar or super-stellar C/O abundance ratios \citep{madhusudhan2014}.
For ultra-hot Jupiters, atmospheric abundances of refractory elements
like magnesium, silicon, and iron also can constrain formation location
\citep[e.g.,][]{lothringer2021}.

It has also been argued that the abundances in giant
planet envelopes depend critically on the assumptions
made regarding the refractory composition of the inner disk
\citep{mordasini2016,cridland2019a,cridland2020}.  Compositions
inherited from the interstellar medium but including the destruction
of carbon grains in the inner disk by oxidizing reactions on grain
surfaces require $\text{C/O} < 1$, both inside and outside the H$_{2}$O
ice line.  The reason is that the accretion of oxygen in the form of
silicate planetesimals in the former case and H$_{2}$O-rich material
in the latter case drive C/O below its abundance in disk gas at a giant
planet's formation location.  In contrast, the assumptions that solids
form along the condensation sequence as a disk cools can still produce
giant planet envelopes with $\text{C/O} > 1$.

All of these analyses rely on two assumptions: (1) that the envelope of
a young giant planet stays well mixed during its formation even though
most metals are accreted before most gas and (2) that a mature giant
planet's atmosphere has a similar composition to the average composition
of its envelope at the end of the planet formation process.  While fully
convective and therefore well-mixed envelopes are the usual outcome
of giant planet formation models, compositional gradients can quench
convection \citep[e.g.,][]{leconte2012}.  In this case, the abundances
in a giant planet's envelope will only reflect the abundances of the gas
it accreted during runaway accretion \citep{thiabaud2015}.  Despite all
of these complications, one robust prediction of giant planet formation
models in dynamically evolving disks is that the metal abundances of
giant planets with $M_{\text{p}} \lesssim 2~M_{\text{Jup}}$ are dominated
by the accretion of planetesimals after envelope accretion.  On the
other hand, the metal abundances of giant planets with $M_{\text{p}}
\gtrsim 2~M_{\text{Jup}}$ are dominated by envelope accretion itself
\citep{mordasini2014,mordasini2016,cridland2019b}.

In short, the interpretation of giant planet atmospheric carbon, oxygen,
and C/O abundances is far from simple.  Moreover, giant planet atmospheric
abundance ratios can only be meaningfully interpreted relative to the
mean compositions of their parent protoplanetary disk.  Because the
protoplanetary disks that formed the observed giant planets disappeared
long ago, the only way to reveal the mean compositions of those disks is
to use the photospheric abundances of their host stars.  During the era of
giant planet formation, the star growing at the center of a protoplanetary
disk has already accreted 99\% of the material that ever passed through
its disk.  As a result, the photospheric abundances of solar-type
planet-host stars are an excellent proxy for mean protoplanetary disk
abundances.  The implication is that accurate and precise host star
elemental abundances for the same elements observed in giant planet
atmospheres are critically needed to achieve the full potential of giant
planet atmospheric abundance inferences as planet formation constraints.

In this article, we infer photospheric and fundamental stellar parameters
as well as individual elemental abundances---including carbon and
oxygen---for the hot Jupiter host star WASP-77 A.  Atmospheric carbon
and oxygen abundances for the hot Jupiter WASP-77 A b were recently
published by \citet{line2021}, and the carbon and oxygen abundances we
infer for WASP-77 A qualitatively change the interpretation of WASP-77
A b's atmospheric abundances.  We describe in Section \ref{data} the
high-resolution optical spectrum we collected for WASP-77 A.  We then
infer stellar parameters from that spectrum and all available astrometric
and photometric data in Section \ref{stellar_param}.  We derive the
individual elemental abundance in WASP-77 A's photosphere in Section
\ref{elem_abund}.  We infer updated parameters for the planet WASP-77 A
b that are self consistent with the stellar parameters we inferred for
its host star in Section \ref{planet_param}.  We review our results
and their implications in Section \ref{discussion}.  We conclude by
summarizing our findings in Section \ref{conclusion}.

\section{Data}\label{data}

The hot Jupiter host star WASP-77 A has right ascension $\alpha$ = 
02 28 37.23, declination $\delta$ = --07 03 38.4, and is also known as 
BD-07 436 A, TYC 4697-201-1, TOI-398, and Gaia EDR3 5178405479961698048.
We observed it from Apache Point Observatory (APO) with the Astrophysical
Research Consortium (ARC) 3.5-m Telescope and its ARC Echelle Spectrograph
\citep[ARCES;][]{wang2003} for 750 seconds starting at UTC 2021 February
23 01:42:55.  We used the standard 1\farcs6 x 3\farcs2 slit, yielding a
spectrum between 320 nm and 1000 nm with spectral resolution $R \approx
31,\!500$.  We collected all calibration data (e.g., bias, quartz flat
field, and ThAr lamp frames) in the late afternoon before the start of
our observations.  We reduced the raw spectra and calibration frames
using a customized version of the \texttt{CERES} echelle spectrograph
data reduction package\footnote{\url{https://github.com/rabrahm/ceres}}
\citep{brahm2017}.  We cross-correlated each order with a theoretical
solar spectrum and averaged the result to calculate the barycentric radial
velocity $v_{r} = 2 \pm 1$ km s$^{-1}$ consistent with the High Accuracy
Radial velocity Planet Searcher (HARPS)-based radial velocity presented
in \citet{maxted2013}.  We then used our radial velocity measurement to
place the normalized spectrum in its rest frame.  Because WASP-77 A is a
solar-type star with close-to-solar photospheric stellar parameters, we
used \texttt{iSpec}\footnote{\url{https://www.blancocuaresma.com/s/iSpec}}
\citep{blanco-cuaresma2014,blanco-cuaresma2019} and a solar template
spectrum to continuum normalize the individual orders.  Our reduced,
continuum-normalized one-dimensional (1D) spectrum has a signal-to-noise
ratio $\text{S/N} \gtrsim 100$ pixel$^{-1}$ at 500 nm and $\text{S/N}
\gtrsim 150$ pixel$^{-1}$ at 620 nm.

\section{Stellar Parameters}\label{stellar_param}

We derive photospheric and fundamental stellar parameters for WASP-77 A
using the algorithm described in \citet{reggiani2020,reggiani2021} that
makes use of both the classical spectroscopy-only approach\footnote{The
classical spectroscopy-only approach to photospheric stellar parameter
estimation involves simultaneously minimizing for individual line-based
iron abundance inferences the difference between \ion{Fe}{1} \&
\ion{Fe}{2}-based abundances as well as their dependencies on transition
excitation potential and measured reduced equivalent width.} and
isochrones to infer accurate, precise, and self-consistent photospheric
and fundamental stellar parameters.  Isochrones are especially useful
for effective temperature $T_{\text{eff}}$ inferences in this case, as
high-quality multiwavelength photometry from the ultraviolet to the red
optical are available from Data Release (DR2) of the SkyMapper Southern
Survey (SMSS) and Gaia DR2.  Similarly, the Gaia EDR3 parallax-based
distance of WASP-77 A makes the calculation of surface gravity $\log{g}$
via isochrones straightforward.  With both $T_{\text{eff}}$ and $\log{g}$
available via isochrones, the equivalent widths of iron lines can
be used to self-consistently determine metallicity $[\text{Fe/H}]$
and microturbulence $\xi$ by minimizing the dependence of individual
line-based iron abundance inferences on reduced equivalent width.

The inputs to our photospheric and fundamental stellar parameter
inference include the equivalent widths of \ion{Fe}{1} and \ion{Fe}{2}
atomic absorption lines, multiwavelength photometry, a Gaia parallax,
and an extinction estimate.  Using atomic absorption line data from
\cite{galarza2019} for lines from \citet{melendez2014} found to be
insensitive to stellar activity, we first measure these equivalent
widths by fitting Gaussian profiles with the \texttt{splot} task in
\texttt{IRAF} to our continuum-normalized spectrum.  Whenever necessary,
we use the \texttt{deblend} task to disentangle absorption lines from
adjacent spectral features.  We gather $u$, $v$, $g$, $r$, and $z$
photometry and their uncertainties from SMSS DR2 \citep{onken2019}
as well as $G$ photometry and its uncertainty from Gaia DR2
\citep{gaia2016,gaia2018,arenou2018,evans2018,hambly2018,riello2018}.
We use a Gaia EDR3 parallax and its uncertainty
\citep{gaia2021,fabricius2021,lindegren2021a,lindegren2021b,torra2021}
as well as an extinction $A_V$ inference based on three-dimensional
(3D) maps of extinction in the solar neighborhood from the STructuring
by Inversion the Local Interstellar Medium (Stilism) program
\citep{lallement2014,lallement2018,capitanio2017}.

We assume \citet{asplund2021} solar abundances and use these inputs
to infer photospheric and fundamental stellar parameters using the
following steps.
\begin{enumerate}
\item
We use 1D plane-parallel solar-composition ATLAS9 model atmospheres
\citep{castelli2004}, the 2019 version of the \texttt{MOOG}
radiative transfer code \citep{sneden1973}, and the \texttt{q$^2$}
\texttt{MOOG} wrapper\footnote{\url{https://github.com/astroChasqui/q2}}
\citep{ramirez2014} to derive an initial set of photospheric stellar
parameters $T_{\text{eff}}$, $\log{g}$, $[\text{Fe/H}]$, and $\xi$
using the classical spectroscopy-only approach.
\item
We then use the \texttt{isochrones}
package\footnote{\url{https://github.com/timothydmorton/isochrones}}
\citep{morton2015} to fit the MESA Isochrones and Stellar Tracks
\cite[MIST;][]{dotter2016,choi2016,paxton2011,paxton2013,paxton2015,paxton2018,paxton2019}
library to our photospheric stellar parameters as well as our
input multiwavelength photometry, parallax, and extinction data using
\texttt{MultiNest}\footnote{\url{https://ccpforge.cse.rl.ac.uk/gf/project/multinest/}}
\citep{feroz2008,feroz2009,feroz2019} via \texttt{PyMultinest}
\citep{buchner2014}.  We restricted the MIST library to extinctions in
the range $0~\text{mag} \leq A_{V} \leq 0.05~\text{mag}$ based on the
maximum plausible $A_{V}$ value suggested by Stilism.  This produces a
new set of photospheric and fundamental stellar parameter posteriors that
are both self-consistent and physically consistent with stellar evolution.
\item
We next impose the posterior-median $T_{\text{eff}}$ and $\log{g}$
inferred in step 2 on our grid of model atmospheres and minimize the
dependence of individual line-based iron abundance inferences on reduced
equivalent width to derive model atmosphere $[\text{Fe/H}]_{\text{atmo}}$
and $\xi$ values consistent with our measured \ion{Fe}{1} \& \ion{Fe}{2}
equivalent widths and our \texttt{isochrones} inferred $T_{\text{eff}}$
\& $\log{g}$.
\item
We then use the model atmosphere selected in step 3 to
calculate $[\text{Fe/H}]$ as the average of all $n_{\text{Fe}} =
n_{\text{\ion{Fe}{1}}} + n_{\text{\ion{Fe}{2}}}$ equivalent width-based
iron abundance inferences for individual \ion{Fe}{1} \& \ion{Fe}{2}
lines.  We take the uncertainty of our $[\text{Fe/H}]$ inference as the
standard deviation of the individual line-based abundance inferences
$\sigma_{[\text{Fe/H}]}'$ divided by $\sqrt{n_{\text{Fe}}}$.
\item
We next check if the $[\text{Fe/H}]$ inferred in step 4 agrees to two
decimal places with $[\text{Fe/H}]_{\text{atmo}}$.  If so, we proceed
to step 6.  If not, we replace $[\text{Fe/H}]_{\text{atmo}}$ with
$[\text{Fe/H}]$ and repeat steps 3 to 5 until agreement is achieved.
\item
We then repeat steps 2 to 5 until the metallicities inferred from
both the \texttt{isochrones} analysis and the reduced equivalent width
balance approach are consistent within their uncertainties (typically
a few iterations).
\end{enumerate}

\begin{figure*}
\plotone{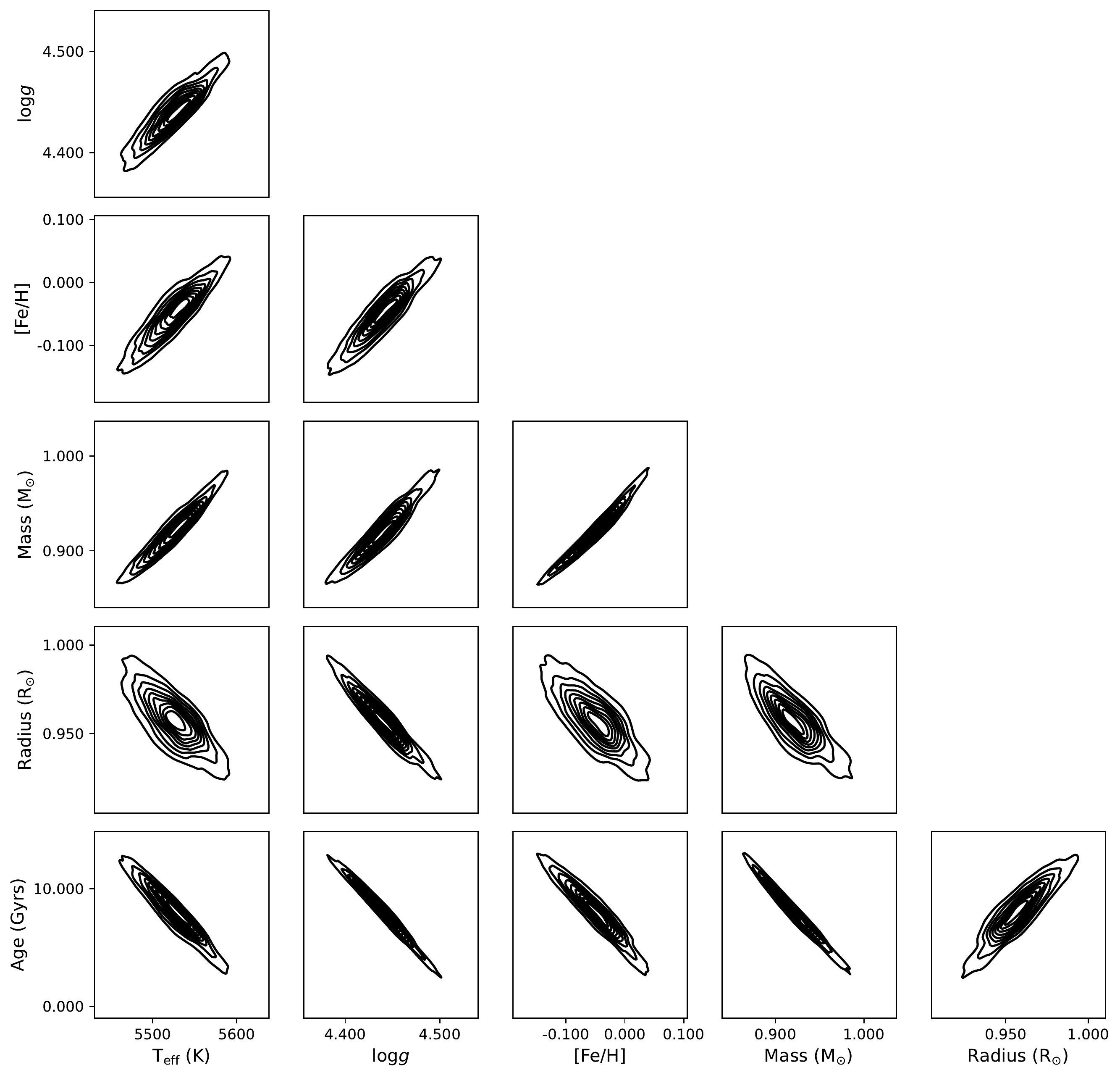}
\caption{Samples from photospheric and fundamental stellar parameter
posteriors from our Bayesian analyses of astrometric, photometric,
and spectroscopic data  for WASP-77 A.  We emphasize that our adopted
$[\text{Fe/H}]$ comes from averaging individual equivalent width-based
iron abundance inferences as described above and not directly from the
isochrones-inferred metallicity plotted here.\label{wasp77a_corner}}
\end{figure*}

We use a Monte Carlo simulation to derive the final values and
uncertainties in our adopted $[\text{Fe/H}]$ and $\xi$ values due to
the uncertainties in our adopted $T_{\text{eff}}$ and $\log{g}$.
\begin{enumerate}
\item
We randomly sample a self-consistent pair of $T_{\text{eff}}$ and
$\log{g}$ from our converged \texttt{isochrones} posteriors described
above and calculate the values of $[\text{Fe/H}]_{\text{atmo}}$ and
$\xi$ that produce the best reduced equivalent width balance given our
\ion{Fe}{1} \& \ion{Fe}{2} equivalent width measurements.
\item
We use the model atmosphere selected in step 1 to calculate the average
of all $n_{\text{Fe}} = n_{\text{\ion{Fe}{1}}} + n_{\text{\ion{Fe}{2}}}$
individual equivalent width-based iron abundance inferences and save
the resulting metallicity of each iteration.
\item
We repeat steps 1 and 2 200 times and adopt as our final photospheric
stellar parameters the (16,50,84) percentiles of the 200 self-consistent
sets of $T_{\text{eff}}$, $\log{g}$, $[\text{Fe/H}]$, and $\xi$ produced
in this way.
\end{enumerate}
We find good agreement between these photospheric stellar parameter
derived from our Monte Carlo simulation and those resulting from a
single iteration of reduced equivalent width balance using the median
$T_{\text{eff}}$ and $\log{g}$ from the posteriors produced by our
converged analysis.  We report our adopted photospheric and fundamental
stellar parameters in Table \ref{stellar_param_table} and plot samples
from their posteriors in Figure \ref{wasp77a_corner}.  All of the
uncertainties quoted in Table \ref{stellar_param_table} include random
uncertainties only.  That is, they are uncertainties derived under the
unlikely assumption that the MIST isochrone grid we use in our analyses
perfectly reproduces all stellar properties.

We use Transiting Exoplanet Survey Satellite \citep[TESS
-][]{ricker2015} data from Sectors 4 and 31 to measure the
rotational period of WASP-77 A.  We used the \texttt{eleanor}
software framework\footnote{\url{https://adina.feinste.in/eleanor/}}
\citep{feinstein2019} to generate light curves for WASP-77 A
based on Sector 4 and 31 TESS full-frame image data.  We masked
data points coincident with transits of WASP-77 A b or that had
non-zero data quality flags.  We used the \texttt{lightkurve}
package\footnote{\url{https://docs.lightkurve.org/}}
\citep{lightkurve2018} to linearly flatten the light curves across a
single-orbit window to preserve variability while removing orbit-to-orbit
trends.  After normalizing each light curve and eliminating 4-$\sigma$
outliers, we followed \citet{healy2020} and \citet{healy2021} to use both
a stellar light curve's autocorrelation function (ACF) and its periodogram
to infer stellar rotation period.  For both the combined light curve
and each sector individually, we used the first peak of the ACF and its
half-width at half-maximum to estimate the rotation period of WASP-77
A and its uncertainty.  In all cases, the ACFs indicate statistically
consistent periods.  The ACF-inferred rotation periods are consistent
with the periodogram peaks for both the Sector 4 and the Sectors 4+31
analyses, though for the Sector 31 analysis the periodogram peak suggests
a rotation period a factor of two higher than the ACF.  Based on the
combined two-sector light curve, the first peak of the ACF produces our
final estimate of WASP-77 A's rotation period $P_{\text{rot}} = 15.6
\pm 3.5$ reported in Table \ref{stellar_param_table}.  The periodogram
of the combined two-sector light curve is consistent with this period
determination.  We use the the rotational evolution models presented
in \citet{vansaders2016} to roughly infer an age $\tau_{\text{gyro}}
\approx 1.5$ Gyr for WASP-77 A based on this rotation period inference.

\begin{deluxetable}{lcc}
\tablecaption{Adopted Stellar Parameters}\label{stellar_param_table}
\tablewidth{0pt}
\tablehead{
\colhead{Property} & \colhead{Value} & \colhead{Unit}}
\startdata
SkyMapper DR2 $u$ & $12.084\pm0.008$ & AB mag \\
SkyMapper DR2 $v$ & $11.756\pm0.008$ & AB mag \\
SkyMapper DR2 $g$ & $10.443\pm0.034$ & AB mag \\
SkyMapper DR2 $r$ & $10.088\pm0.04$ & AB mag \\
SkyMapper DR2 $z$ & $9.881\pm0.046$ & AB mag \\
Gaia DR2 DR2 $G$   & $10.096\pm0.002$ & Vega mag \\
Gaia EDR3 parallax & $9.459\pm0.056$ & mas \\
\hline
\multicolumn{3}{l}{\textbf{Isochrone-inferred parameters}} \\
Effective temperature $T_{\text{eff}}$ & $5525^{+23}_{-25}$ & K \\
Surface gravity $\log{g}$ & $4.44 \pm 0.02$ & cm s$^{-2}$ \\
Stellar mass $M_{\ast}$ & $0.92 \pm 0.02$ & $M_{\odot}$ \\
Stellar radius $R_{\ast}$ & $0.96^{+0.01}_{-0.02}$ & $R_{\odot}$ \\
Luminosity $L_{\ast}$ & $0.77 \pm 0.01$ & $L_{\odot}$ \\
Isochrone-based age $\tau_{\text{iso}}$ & $7.99^{+1.90}_{-1.71}$ & Gyr \\
\hline
\multicolumn{3}{l}{\textbf{Spectroscopically inferred parameters}} \\
$[\text{Fe/H}]$ &  $-0.05^{+0.02}_{-0.01}$ & \\
$\xi$ & $1.26 \pm 0.02$ & km s$^{-1}$ \\
\hline
\multicolumn{3}{l}{\textbf{Light-curve inferred parameters}} \\
Rotational period $P_{\text{rot}}$ & $15.6 \pm 3.5$ & day \\
Gyrochronology-based age $\tau_{\text{gyro}}$ & 1.5 & Gyr \\
\enddata
\end{deluxetable}

To evaluate the impact any possible systematic uncertainties resulting
from our analysis, we compare the photospheric and fundamental stellar
parameters we infer for WASP-77 A with those reported by other groups for
the same star.  WASP-77 A was initially studied by \citet{maxted2013} in
the paper announcing the discovery of WASP-77 A b.  Photospheric and/or
fundamental parameters for the star based on high-resolution spectroscopy
have since been published by \citet{mortier2013} \citep[updated
in][]{sousa2021} and \citet{kolecki2021}.  \citet{maxted2013}
found $T_{\text{eff}} = 5500 \pm 80$ K, $\log{g} = 4.33 \pm 0.08$,
$[\text{Fe/H}] = 0.00 \pm 0.11$, $M_{\ast} = 1.00 \pm 0.07~M_{\odot}$,
$\tau_{\text{iso}} = 8$ Gyr, and $P_{\text{rot}} = 15.4 \pm 0.4$ days.
\citet{mortier2013} found $T_{\text{eff}} = 5605 \pm 41$ K, $\log{g} =
4.37 \pm 0.09$, $[\text{Fe/H}] = 0.07 \pm 0.03$, and $M_{\ast} = 0.96
\pm 0.07~M_{\odot}$.  Updated parameters from the same group presented
in \citet{sousa2021} found $T_{\text{eff}} = 5595 \pm 16$ K, $\log{g}
= 4.41 \pm 0.03$, $[\text{Fe/H}] = 0.01 \pm 0.01$, and $M_{\ast} =
0.915 \pm 0.004~M_{\odot}$.  \citet{kolecki2021} found $T_{\text{eff}}
= 5660 \pm 30$ K, $\log{g} = 4.49 \pm 0.01$, and $[\text{Fe/H}] = -0.15
\pm 0.06$.  Our photospheric and fundamental stellar parameters agree
well with those presented in \citet{maxted2013} and \citet{mortier2013}
as updated by \citet{sousa2021}.  There is some tension between our
parameters and those reported by \citet{kolecki2021} that we attribute to
their neglect of extinction and their use of Two Micron All Sky Survey
(2MASS) and Wide-field Infrared Survey Explorer (WISE) photometry for
WASP-77 A that were contaminated by its binary companion WASP-77 B.
The excellent agreement between our photospheric and fundamental stellar
parameters for WASP-77 A and those inferred by other groups suggests
that any systematic uncertainties present must be small.

As an additional check we inferred $T_{\text{eff}}$ for WASP-77 A using
the \texttt{colte} code\footnote{\url{https://github.com/casaluca/colte}}
\citep{casagrande2021} that estimates $T_{\text{eff}}$ using a combination
of color--$T_{\text{eff}}$ relations obtained by implementing the InfraRed
Flux Method for Gaia and 2MASS photometry.  As required by \texttt{colte},
we used Gaia EDR3 $G$, $G_{\text{BP}}$, and $G_{\text{RP}}$ plus 2MASS
$J$, $H$, and $K_s$ photometry as input.  We used these data despite the
fact that 2MASS $J$ and $H$ are affected by confusion and 2MASS $K_s$
is affected by a diffraction spike.  We find a \texttt{colte}-based
$T_{\text{eff}} = 5354 \pm 203$ K.  While we suspect its large uncertainty
is due to the poor quality of the required input 2MASS photometry,
it is consistent with our adopted $T_{\text{eff}} = 5525^{+23}_{-25}$ K.

There are significant differences between the ages inferred
for WASP-77 A from isochrones and from gyrochronology.  We find
$\tau_{\text{iso}}$ = $7.99^{+1.90}_{-1.71}$ Gyr in full agreement
with the $\tau_{\text{iso}} = 8$ Gyr inference from the full MCMC
fit to all data from \citet{maxted2013}.  These inferences are also
consistent with the fits produced by five additional isochrones grids
presented in \citet{maxted2013} using the methodology described in
\citet{southworth2012}.  Similarly, if we fit Yonsei-Yale isochrones
\citep{yi2001,yi2003,kim2002,demarque2004} to our $T_{\text{eff}}$ and
$\log{g}$ inferences using \texttt{q$^2$} we find $\tau_{\text{iso}}
= 8.3 \pm 3.5$ Gyr.  On the other hand, \citet{maxted2013} used their
$P_{\text{rot}} = 15.4 \pm 0.4$ days to infer a gyrochronology-based age
$\tau_{\text{gyro}} = 1.0^{+0.5}_{-0.3}$ Gyr using the gyrochronology
relation from \citet{barnes2007}.  Using our TESS-based $P_{\text{rot}}
= 15.6 \pm 3.5$ days we find $\tau_{\text{gyro}} \approx 1.5$ Gyr using
the updated \citet{vansaders2016} models.

We argue that $\tau_{\text{iso}}$ should be preferred
to $\tau_{\text{gyro}}$ for WASP-77 A.  We assume that
WASP-77 B has the same metallicity and extinction we
inferred for WASP-77 A and use the same procedure described
above to fit the MIST isochrone grid to Sloan Digital Sky Survey
\citep[SDSS;][]{fukugita1996,gunn1998,gunn2006,york2000,doi2010,abdurrouf2021}
$u$ and Gaia DR2 $G$ photometry as well as a Gaia EDR3 parallax for
WASP-77 B.  We find consistent ages for WASP-77 A $\tau_{\text{iso}}$
= $7.99^{+1.90}_{-1.71}$ and WASP-77 B $\tau_{\text{iso}}$
= $7.3^{+4.4}_{-4.6}$.  We also note that \citet{salz2015} studied
the X-ray properties of the WASP-77 system and found X-ray-based ages
consistent with our isochrone-based age inferences for both stars.
As we will show in Section \ref{discussion}, the difference between the
isochrone- and gyrochronology-based age inferences results from tidally
mediated angular momentum exchange between the WASP-77 A system's orbital
angular momentum and the rotational angular momentum of WASP-77 A.

It has been proposed that the chemical abundances of solar neighborhood
``solar twin'' stars can be used as ``chemical clocks'' to complement
other age inference techniques.  WASP-77 A is a thin disk star, so
its elemental abundances are the result of thin disk Galactic chemical
evolution.  We use our inferred elemental abundances and the empirical
chemical clocks derived from thin disk solar twins by \citet{spina2018}
to infer two more, independent, age estimates based on $[\text{Y/Mg}]$ and
$[\text{Y/Al}]$ abundance ratios.  According to Equations (4) and (5) from
\citet{spina2018}, WASP-77 A has a $[\text{Y/Mg}]$ chemical clock-based
age $\tau_{\text{cc}}= 8.0 \pm 1.0$ Gyr and a $[\text{Y/Al}]$ chemical
clock-based age $\tau_{\text{cc}}= 6.5 \pm 0.9$ Gyr.  The consistency of
$\tau_{\text{iso}}$ and these $\tau_{\text{cc}}$ estimates supports our
interpretation of the gyrochronology-based age inferences for WASP-77 A.

\section{Elemental Abundances}\label{elem_abund}

To infer the elemental abundances of several $\alpha$, light
odd-$Z$, iron-peak, and neutron-capture elements we first measure
the equivalent widths of atomic absorption lines of \ion{C}{1},
\ion{O}{1}, \ion{Na}{1}, \ion{Mg}{1}, \ion{Al}{1}, \ion{Si}{1},
\ion{K}{1}, \ion{Ca}{1}, \ion{Sc}{2}, \ion{Ti}{1}, \ion{Ti}{2},
\ion{V}{1}, \ion{Cr}{1}, \ion{Fe}{1}, \ion{Fe}{2}, \ion{Ni}{1}, and
\ion{Y}{2} in our continuum-normalized spectrum by fitting Gaussian
profiles with the \texttt{splot} task in \texttt{IRAF}.  We use the
\texttt{deblend} task to disentangle absorption lines from adjacent
spectral features whenever necessary.  We measure an equivalent width for
every absorption line in our line list that could be recognized, taking
into consideration the quality of the spectrum in the vicinity of a line
and the availability of alternative transitions of the same species.
We assume \citet{asplund2021} solar abundances and local thermodynamic
equilibrium (LTE) and use the 1D plane-parallel solar-composition
ATLAS9 model atmospheres and the 2019 version of \texttt{MOOG} to
infer elemental abundances based on each equivalent width measurement.
We report our adopted atomic data, equivalent width measurements, and
individual line-based abundance inferences in Table \ref{measured_ews}.

\begin{deluxetable*}{lccccc}
\tablecaption{Atomic Data, Equivalent-width Measurements, and
Individual-line Abundance Inferences\label{measured_ews}}
\tablewidth{0pt}
\tablehead{
\colhead{Wavelength} & \colhead{Species} &
\colhead{Excitation Potential} & \colhead{log($gf$)} &
\colhead{EW} & \colhead{$\log_\epsilon(\rm{X})$} \\ 
 \colhead{(\AA)} &  & \colhead{(eV)} & & (m\AA) & }
\startdata
$6154.225$ & \ion{Na}{1} & $2.102$ & $-1.547$ & $47.8$ & $6.302$ \\
$6160.747$ & \ion{Na}{1} & $2.104$ & $-1.246$ & $59.4$ & $6.169$ \\
$4571.095$ & \ion{Mg}{1} & $0.000$ & $-5.623$ & $118.3$ & $7.374$ \\
$4730.040$ & \ion{Mg}{1} & $4.340$ & $-2.389$ & $71.6$ & $7.595$ \\
$6319.236$ & \ion{Mg}{1} & $5.108$ & $-2.165$ & $67.0$ & $7.986$ \\
$6318.717$ & \ion{Mg}{1} & $5.108$ & $-1.945$ & $54.4$ & $7.603$ \\
$6696.018$ & \ion{Al}{1} & $3.143$ & $-1.481$ & $49.4$ & $6.453$ \\
$6698.667$ & \ion{Al}{1} & $3.143$ & $-1.782$ & $28.3$ & $6.391$ \\
$7835.309$ & \ion{Al}{1} & $4.021$ & $-0.680$ & $46.7$ & $6.371$ \\
$7836.134$ & \ion{Al}{1} & $4.021$ & $-0.450$ & $66.1$ & $6.376$
\enddata
\tablecomments{This table is published in its entirety in the
machine-readable format.  A portion is shown here for guidance regarding
its form and content.}
\end{deluxetable*}

We report in Table \ref{elem_abundances} our abundance inferences
in three common systems: $A(\text{X})$, $[\text{X/H}]$,
and $[\text{X/Fe}]$.  The abundance $A(\text{X})$ is defined
$A(\text{X})=\log{N_{\text{X}}/N_{\text{H}}} + 12$, the abundance
ratio $[\text{X/H}]$ is defined as $[\text{X/H}] = A(\text{X}) -
A(\text{X})_{\odot}$, and the abundance ratio $[\text{X/Fe}]$ is
defined as $[\text{X/Fe}] = [\text{X/H}] - [\text{Fe/H}]$.  We define
the uncertainty in the abundance ratio $\sigma_{[\text{X/H}]}$ as the
standard deviation of the individual line-based abundance inferences
$\sigma_{[\text{X/H}]}'$ divided by $\sqrt{n_{\text{X}}}$.  We define
the uncertainty $\sigma_{[\text{X/Fe}]}$ as the square root of the sum
of squares of $\sigma_{[\text{X/H}]}$ and $\sigma_{[\text{Fe/H}]}$.

\begin{deluxetable}{lcccccc}
\tablecaption{Elemental Abundances}\label{elem_abundances}
\tablewidth{0pt}
\tablehead{
\colhead{Species} &
\colhead{$A(\text{X})$} &
\colhead{[X/H]} &
\colhead{$\sigma_{\text{[X/H]}}$} & 
\colhead{[X/Fe]} &
\colhead{$\sigma_{[\text{X/Fe}]}$} &
\colhead{$n$}}
\startdata
\multicolumn{6}{l}{\textbf{LTE abundances}} \\
\ion{C}{1} & $8.57$ & $0.11$ & $0.09$ & $0.16$ & $0.11$ & $5$ \\
\ion{O}{1} & $9.03$ & $0.34$ & $0.02$ & $0.39$ & $0.04$ & $3$ \\
\ion{Na}{1} & $6.24$ & $0.02$ & $0.05$ & $0.07$ & $0.07$ & $2$ \\
\ion{Mg}{1} & $7.64$ & $0.09$ & $0.11$ & $0.14$ & $0.13$ & $4$ \\
\ion{Al}{1} & $6.46$ & $0.03$ & $0.04$ & $0.08$ & $0.05$ & $6$ \\
\ion{Si}{1} & $7.58$ & $0.07$ & $0.02$ & $0.12$ & $0.03$ & $13$ \\
\ion{K}{1} & $5.20$ & $0.13$ & $\cdots$ & $0.18$ & $0.06$ & $1$ \\
\ion{Ca}{1} & $6.29$ & $-0.01$ & $0.03$ & $0.04$ & $0.05$ & $11$ \\
\ion{Sc}{2} & $3.04$ & $-0.11$ & $0.03$ & $-0.05$ & $0.06$ & $4$ \\
\ion{Ti}{1} & $4.92$ & $-0.06$ & $0.03$ & $0.00$ & $0.05$ & $14$ \\
\ion{Ti}{2} & $5.00$ & $0.03$ & $0.04$ & $0.08$ & $0.06$ & $10$ \\
\ion{V}{1} & $3.84$ & $-0.06$ & $0.03$ & $-0.01$ & $0.06$ & $8$ \\
\ion{Cr}{1} & $5.51$ & $-0.11$ & $0.05$ & $-0.06$ & $0.06$ & $10$ \\
\ion{Fe}{1} & $7.39$ & $-0.07$ & $0.01$ & $\cdots$ & $\cdots$ & $80$ \\
\ion{Fe}{2} & $7.50$ & $0.04$ & $0.03$ & $\cdots$ & $\cdots$ & $17$ \\
\ion{Ni}{1} & $6.23$ & $0.03$ & $0.02$ & $0.08$ & $0.03$ & $17$ \\
\ion{Y}{2} & $2.14$ & $-0.08$ & $0.05$ & $-0.02$ & $0.08$ & $5$ \\
\hline
\multicolumn{7}{l}{\textbf{1D non-LTE abundances}} \\
\ion{C}{1} & $8.42$ & $-0.04$ & $\cdots$ & $-0.03$ & $\cdots$ & $1$ \\
\ion{O}{1} & $8.93$ & $0.24$ & $\cdots$ &$0.25$ & $\cdots$ & 3\tablenotemark{a} \\
\ion{Na}{1} & $6.16$ & $-0.06$ & $0.10$ &$-0.05$ & $0.10$ & $2$ \\
\ion{Al}{1} & $6.35$ & $-0.08$ & $0.01$ &$-0.07$ & $0.01$ & $2$ \\
\ion{Si}{1} & $7.56$ & $0.05$ & $0.03$ &$0.06$ & $0.03$ & $13$ \\
\ion{K}{1} & $4.89$ & $-0.18$ & $\cdots$ & $-0.17$ & $\cdots$ & $1$ \\
\ion{Ca}{1} & $6.07$ & $-0.23$ & $0.01$ & $-0.22$ & $0.01$ & $2$ \\
\ion{Fe}{1} & $7.42$ & $-0.04$ & $0.01$ & $\cdots$ & $\cdots$ & $80$ \\
\ion{Fe}{2} & $7.53$ & $0.07$ & $0.03$ & $\cdots$ & $\cdots$ & $17$ \\
\hline
\multicolumn{6}{l}{\textbf{3D non-LTE abundances}} \\
\ion{C}{1} & $8.56$ & $0.10$ & $0.09$ & $0.15$ & $0.09$ & $5$ \\
\ion{O}{1} & $8.92$ & $0.23$ & $0.02$ & $0.28$ & $0.02$ & $3$ \\
\hline
\multicolumn{7}{l}{\textbf{Additional abundance ratios of interest}} \\
\multicolumn{7}{l}{$[\text{Fe/H}]_{\text{1D non-LTE}} = -0.01 \pm 0.01$} \\
\multicolumn{7}{l}{$[\text{(C+O)/H}]_{\text{3D non-LTE}} = 0.33 \pm 0.09$} \\
\multicolumn{7}{l}{$[\text{C/O}]_{\text{3D non-LTE}} = -0.13 \pm 0.09$} \\
\multicolumn{7}{l}{$\text{C/O}_{\text{3D non-LTE}} = 0.44^{+0.07}_{-0.08}$}
\enddata
\tablenotetext{a}{1D non-LTE corrections from \citet{amarsi2020} for
the three lines of the oxygen triplet are limited to a mean correction.}
\end{deluxetable}

When possible, we update our elemental abundances derived under the
assumptions of LTE to account for departures from LTE (i.e., non-LTE
corrections) by linearly interpolating published grids of non-LTE
corrections using \texttt{scipy} \citep{virtanen2020}.  We make
use of 1D non-LTE corrections for carbon \citep{amarsi2020}, oxygen
\citep{amarsi2020}, sodium \citep{lind2011}, aluminum \citep{amarsi2020},
silicon \citep{amarsi2017}, potassium \citep{reggiani2019}, calcium
\citep{amarsi2020}, and iron \citep{amarsi2016}.  We also make use
of 3D non-LTE corrections for carbon and oxygen \citep{amarsi2019}.
We present these abundances corrected for departures from LTE in Table
\ref{elem_abundances}.

Our 1D non-LTE abundance corrections for carbon and oxygen come
from \citet{amarsi2020} and are limited to one carbon line and a mean
correction for the three lines of the oxygen triplet.  We therefore cannot
calculate individual line-based abundance dispersions for our 1D non-LTE
corrected carbon and oxygen abundance inferences, so our uncertainties
might be underestimated.  On the other hand, our 3D non-LTE carbon and
oxygen abundance corrections come from \cite{amarsi2019} and are available
for all the lines in our line list.  We do find good agreement between
our 1D non-LTE and 3D non-LTE corrected carbon abundances for the one
line with corrections in both sources.  Because both our 1D non-LTE and
3D non-LTE corrected carbon and oxygen abundance inferences for WASP-77
A indicate that it has a sub-solar C/O ratio, our preference for 3D
non-LTE corrected abundances does not affect the constraints we derive
in Section \ref{discussion} on the formation of WASP-77 A b.

Elemental abundances for WASP-77 A have also been published in
\citet{maxted2013} and \citet{kolecki2021}.  All of our elemental
abundance inferences are consistent with those presented in
\citet{maxted2013}.  Our abundance inferences for carbon, sodium,
magnesium, and silicon are consistent with those presented in
\citet{kolecki2021}, though our oxygen and potassium abundances are
significantly higher than those presented in that paper.  We are able
to reproduce the oxygen abundance reported in \citet{kolecki2021} using
their photospheric stellar parameters, our equivalent width measurements,
and 3D non-LTE abundance corrections from \cite{amarsi2019}.  We are
therefore confident that the tension between our oxygen abundances
results from the photospheric stellar parameters \citet{kolecki2021}
derived for WASP-77 A that are hotter and more metal-poor than those
derived by us, \citet{maxted2013}, and \citet{mortier2013} as updated
by \citet{sousa2021}.

\section{Updated Planetary Parameters}\label{planet_param}

\begin{figure*}
\plotone{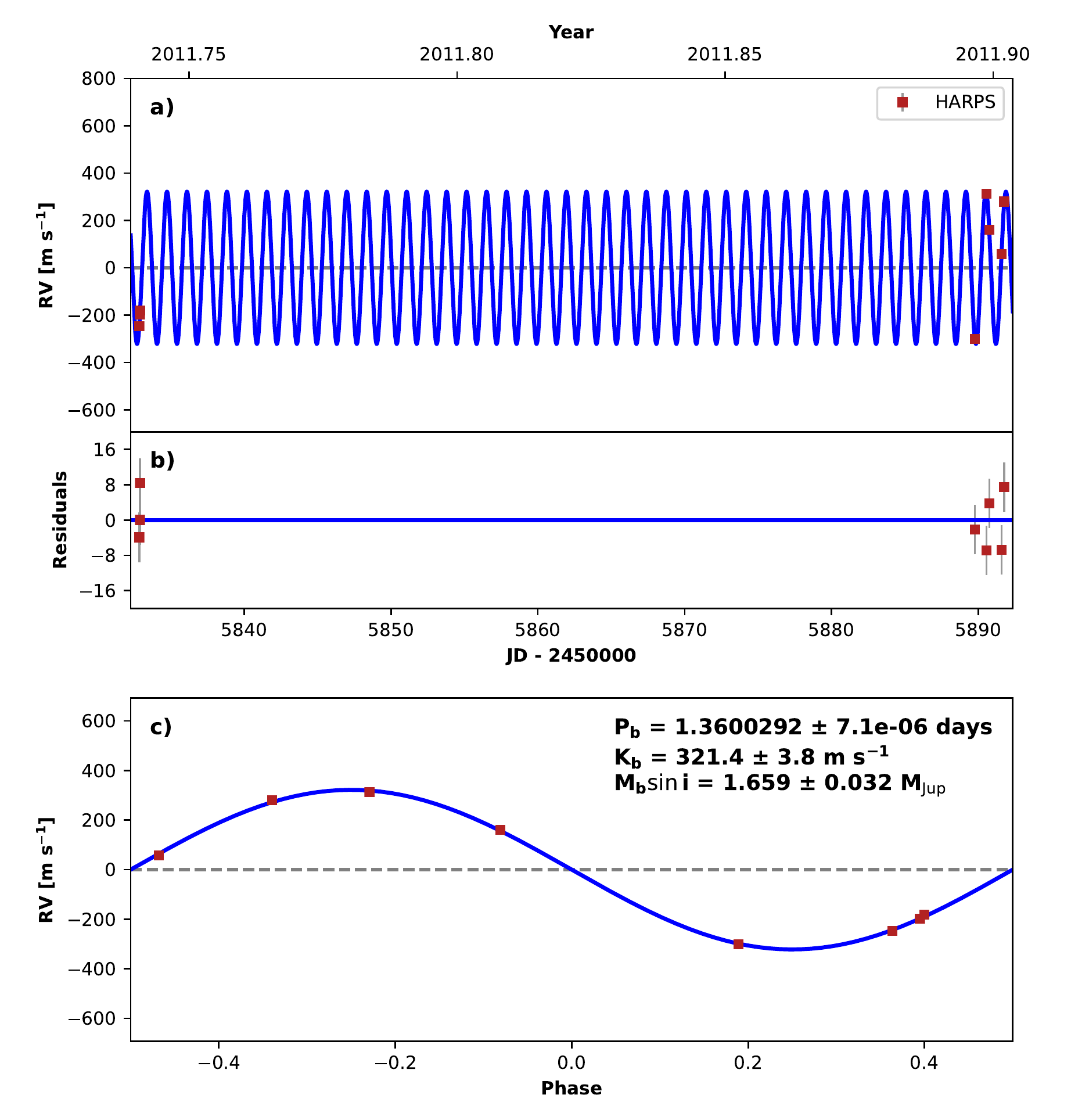}
\caption{\texttt{RadVel} fit of a Keplerian orbit to the WASP-77 A
differential radial velocities from \citet{maxted2013} as updated in
\citet{trifonov2020}.  We fix the orbital period $P = 1.3600306 \pm
0.0000012$ to the value from \citet{turner2016} and the eccentricity to
zero as expected for short-period hot Jupiter orbiting a mature solar-type
star.  Using the stellar mass we inferred from our self-consistent
analysis of astrometric, photometric, and spectroscopic data for WASP-77
A $M_{\ast} = 0.92 \pm 0.02~M_{\odot}$ we find that WASP-77 A b's mass
is $M_{\text{p}} = 1.66 \pm 0.03~M_{\text{Jup}}$.\label{radvel_fit}}
\end{figure*}

We make use of our accurate, precise, and self-consistent stellar
mass and radius inferences for WASP-77 A to update the parameters
of the planet WASP-77 A b.  Taking the transit depth $\delta =
(R_{\text{p}}/R_{\ast})^2 = 0.01693\pm0.00017$ from \citet{maxted2013},
our stellar radius $R_{\ast} = 0.96^{+0.01}_{-0.02}~R_{\odot}$ implies
a planet radius $R_{\text{p}} = 0.125^{+0.001}_{-0.004}~R_{\odot} =
1.22^{+0.01}_{-0.04}~R_{\text{Jup}}$ in agreement with the $R_{\text{p}}
= 1.21 \pm 0.02~R_{\text{Jup}}$ value inferred by \citet{maxted2013}.
We fit the HARPS radial velocities from \citet{maxted2013}
as updated in \citet{trifonov2020} using the \texttt{RadVel}
package\footnote{\url{https://radvel.readthedocs.io/en/latest/}}
\citep{fulton2017,fulton2018}.  We fix the orbital period $P = 1.3600306
\pm 0.0000012$ days to the value from \citet{turner2016}.  We fix the
eccentricity $e = 0$ as expected for such a short-period hot Jupiter
and find a planet mass $M_{\text{p}} = 1.66 \pm 0.03~M_{\text{Jup}}$,
marginally lower than those presented in \citet{maxted2013} and
\citet{bonomo2017}.  We plot our fit to the Doppler data in Figure
\ref{radvel_fit} and summarize our updated planet parameters in Table
\ref{planet_param_table}.

\begin{deluxetable}{lrr}
\tablecaption{Updated Planetary Parameters\label{planet_param_table}}
\tablehead{\colhead{Parameter} & 
\colhead{Credible Interval} & 
\colhead{Units}
}
\startdata
\multicolumn{3}{l}{\textbf{Fixed values}} \\
Orbital period $P_{\text{orb}}$ & $1.3600306 \pm 0.0000012$ & days \\
Eccentricity $e$ & 0.00 &  \\
\hline
\multicolumn{3}{l}{\textbf{Fit values}} \\
Argument of periapse $\omega$ & $177 \pm 120$ & degrees \\ 
Doppler semiamplitude $K_{\ast}$ & $321.4^{+3.9}_{-3.8}$ & m s$^{-1}$ \\
Planet mass $M_{\text{p}}$ & $1.66 \pm 0.03$ & $M_{\text{Jup}}$ \\
Planet radius $R_{\text{p}}$ & $1.22^{+0.01}_{-0.04}$ & $R_{\text{Jup}}$ \\
Planet density $\rho_{\text{p}}$ & $1.14^{+0.12}_{-0.10}$ & g cm$^{-3}$ \\
Semimajor axis $a$ & $0.02338 \pm 0.00017$ & AU \\
\enddata
\end{deluxetable}

\section{Discussion}\label{discussion}

\subsection{The Atmospheric Abundances of WASP-77 A b Indicate it Formed 
Beyond the H$_{2}$O Ice Line}

For the photosphere of WASP-77 A and therefore the protoplanetary disk
in which WASP-77 A b formed, we find  super-solar carbon and oxygen
abundances with a sub-solar C/O abundance ratio\footnote{According to
\citet{asplund2021}, the solar C/O abundance ratio is $\text{C/O} =
0.59^{+0.08}_{-0.06}$.}.  We infer $[\text{C/H}]_{\text{3D non-LTE}} =
0.10 \pm 0.09$ and $[\text{O/H}]_{\text{3D non-LTE}} = 0.23 \pm 0.02$,
implying $[\text{(C+O)/H}]_{\text{3D non-LTE}} = \left[A(\text{C})_{\ast}
- A(\text{C})_{\odot}\right] + \left[A(\text{O})_{\ast} -
A(\text{O})_{\odot}\right] = 0.33 \pm 0.09$ and C/O abundance ratios
$[\text{C/O}]_{\text{3D non-LTE}} = [\text{C/H}] - [\text{O/H}] =
-0.13 \pm 0.09$ and $\text{C/O}_{\text{3D non-LTE}} = 10^{A(\text{C})
- A(\text{O)}} = 0.44^{+0.07}_{-0.08}$.

For the atmosphere of the hot Jupiter WASP-77 A b, \citet{line2021}
found $[\text{C/H}] = -0.46^{+0.17}_{-0.16}$, $[\text{O/H}] =
-0.49^{+0.14}_{-0.12}$, $[\text{(C+O)/H}] = -0.48^{+0.15}_{-0.13}$,
and $\text{C/O} = 0.59 \pm 0.08$.  Those authors also considered the
possibility that cloud condensates like MgSiO$_{3}$ and Mg$_{2}$SiO$_{4}$
``rain out'' on WASP-77 A b's night side and remove oxygen from its
atmosphere.  That process would enhance the gas-phase C/O ratio of
WASP-77 A b.  When accounting for this possibility, \citet{line2021}
found $[\text{(C+O)/H}] = -0.41 \pm 0.14$ and $\text{C/O} = 0.46 \pm
0.08$ for the atmosphere of WASP-77 A b.  \citet{line2021} assumed
that because of its near solar metallicity the star WASP-77 A has a
solar C/O abundance ratio.  Based on that assumption, those authors
asserted that the atmosphere of the planet WASP-77 A b has sub-stellar
$[\text{C/H}]$ and $[\text{O/H}]$ plus a stellar C/O abundance ratio.
They suggested that WASP-77 A b accreted its envelope interior to its
parent protoplanetary disk's H$_{2}$O ice line from carbon-depleted gas
with little subsequent planetesimal accretion or core erosion.

We find that carbon and oxygen are 3.6 times and 5.2 times less
abundant in the atmosphere of WASP-77 A b than in the photosphere
of WASP-77 A.  Our abundance inferences for WASP-77 A increase the
statistical significance of the \citet{line2021} observation that WASP-77
A b's atmosphere has sub-stellar carbon and oxygen abundances, from 2.6
$\sigma$ to 2.9 $\sigma$ for carbon and from 3.4 $\sigma$ to 5.1 $\sigma$
for oxygen.  Our C/O abundance ratio inference for WASP-77 A qualitatively
changes the status of WASP-77 A b's atmospheric C/O abundance ratio from
stellar to super-stellar by a factor of 1.4 at the 1.4-$\sigma$ level.

As we argued in Section \ref{intro}, it is not trivial to connect
a giant planet's atmospheric carbon and oxygen abundances to its
formation location in its parent protoplanetary disk.  Nevertheless,
WASP-77 A b's significantly sub-stellar carbon and oxygen abundances
combined with a super-stellar C/O abundance ratio is consistent
with formation beyond its parent protoplanetary disk's H$_{2}$O ice
line \citep[e.g.,][]{oberg2011}, possibly from pebble accretion
followed by high-eccentricity migration after disk dissipation
\citep[e.g.,][]{madhusudhan2014,madhusudhan2017,booth2017}.  The
possibility that some of the oxygen in WASP-77 A b's atmosphere has
rained-out in the form of condensates means that the bulk C/O ratio of
the planet may indeed match the stellar abundance.

\begin{figure*}
\plottwo{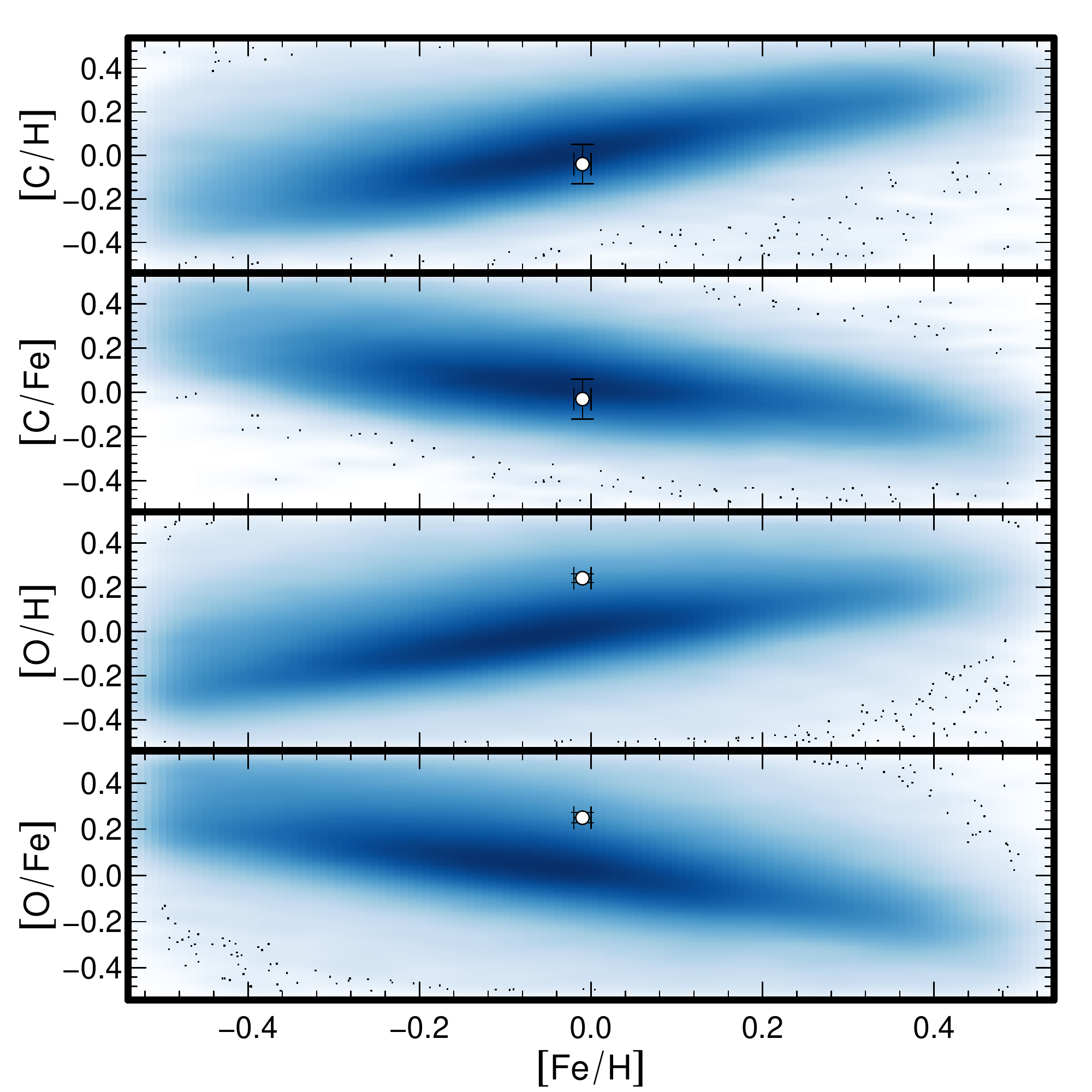}{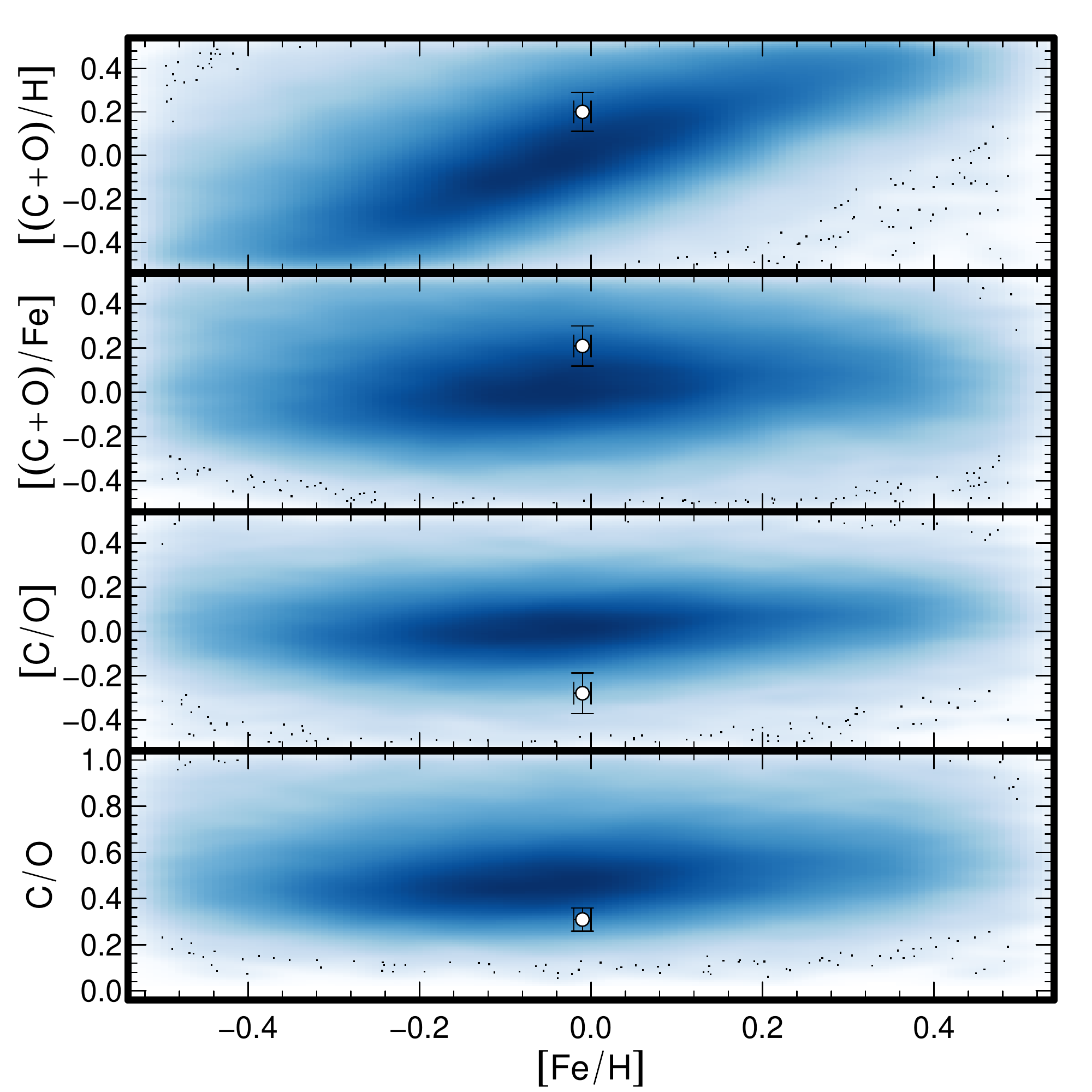}
\caption{1D non-LTE carbon and oxygen abundances of WASP-77 A
compared to the same 1D non-LTE abundances for solar neighborhood
dwarfs from GALAH DR3 \citep{buder2021}.  In each kernel density
estimator-smoothed scatter plot, darker blue regions correspond
to more densely populated regions of abundance space.  We plot
outliers as individual dots.  We plot our abundance inferences for
WASP-77 A's as black-bordered white points.  Left: $[\text{C/H}]$,
$[\text{C/Fe}]$, $[\text{O/H}]$, and $[\text{O/Fe}]$.  Right:
$[\text{(C+O)/H}]$, $[\text{(C+O)/Fe}]$, $[\text{C/O}]$, and C/O.
In the range $-0.1 \leq [\text{Fe/H}] \leq 0$, WASP-77 A is consistent
with the GALAH-inferred carbon ($[\text{C/Fe}] = 0.05^{+0.12}_{-0.10}$),
oxygen ($[\text{O/Fe}] = 0.06^{+0.14}_{-0.10}$), and C/O abundance ratio
distributions ($[\text{C/O}] = 0.01^{+0.11}_{-0.12}$ and $\text{C/O} =
0.49^{+0.15}_{-0.12}$).  In the metallicity range $-0.5 \leq [\text{Fe/H}]
\leq 0.5$ appropriate for virtually all known exoplanet host stars,
GALAH DR3 finds that 95\% of solar neighborhood dwarf stars have
$-0.15 \leq [\text{C/Fe}] \leq 0.32$, $-0.19 \leq [\text{O/Fe}] \leq
0.43$, $-0.26 \leq [\text{C/O}] \leq 0.29$, and $0.27 \leq \text{C/O}
\leq 0.94$.  These broad distributions emphasize that the assumption of
solar abundance ratios of carbon and oxygen for exoplanet host stars is
not always appropriate.\label{fig01}}
\end{figure*}

The qualitative change in the interpretation of WASP-77 A b's atmospheric
carbon and oxygen abundances brought about by our inference of WASP-77
A's non-solar $[\text{C/H}]$, $[\text{O/H}]$, and C/O abundance
ratio emphasizes the importance of stellar abundance inferences for
the interpretation of giant planet atmospheric abundances.  Indeed,
even though WASP-77 A has a solar metallicity $[\text{Fe/H}]_{\text{1D
non-LTE}} = -0.01 \pm 0.01$ it has non-solar carbon and oxygen abundances.
To illustrate this point, we plot in Figure \ref{fig01} the 1D non-LTE
carbon and oxygen abundances of solar neighborhood dwarf stars (i.e.,
$\log{g} \geq 4$) over the 1D non-LTE correct metallicity range
characteristic of virtually all known planet host stars $-0.5 \leq
[\text{Fe/H}] \leq 0.5$ from the third data release of the Galactic
Archaeology with HERMES (GALAH) survey \citep{buder2021}.  Figure
\ref{fig01} shows that WASP-77 A is fully consistent with the solar
neighborhood carbon and oxygen abundance distributions.  It also shows
that the solar neighborhood carbon, oxygen, and C/O abundance ratios
are much broader than usually considered in planet formation models,
with 95\% of solar neighborhood dwarf stars in the ranges $-0.15 \leq
[\text{C/Fe}] \leq 0.32$, $-0.19 \leq [\text{O/Fe}] \leq 0.43$, $-0.26
\leq [\text{C/O}] \leq 0.29$, and $0.27 \leq \text{C/O} \leq 0.94$.

If a planet host star's carbon and oxygen abundances are unknown,
then a planet with a sub-stellar but super-solar C/O abundance ratio
would be mischaracterized as carbon rich.  Likewise, a planet with a
super-stellar but sub-solar C/O abundance ratio would be mischaracterized
as carbon poor.  We therefore recommend that all future studies of
exoplanet atmospheric abundances also infer stellar photospheric and
fundamental parameters that are self-consistent and physically consistent
with stellar evolution.  Those photospheric stellar parameters should
be used in all elemental abundance inferences, and those fundamental
stellar parameters should be used to recalculate exoplanet parameters
(e.g, $M_{\text{p}}$, $R_{\text{p}}$, $\log{g}_{\text{p}}$, etc) using
observables like Doppler velocities and light curves.  In that way, all of
the parameters necessary to characterize an exoplanet system can be made
self-consistent to eliminate the small but possibly important systematic
uncertainties that can effect an exoplanet atmosphere abundance inference.
If the photospheric stellar parameters that produce carbon and oxygen
abundances are not used to self consistently infer (1) stellar masses and
radii and (2) planet parameters using observables, then those photospheric
abundances are not suitable for exoplanet atmospheric characterization
\citep[e.g.,][]{kolecki2021}.

\subsection{Angular Momentum Exchange Explains the Tension Between 
Isochrone- and Gyrochronology-based Ages for WASP-77 A}\label{gyro_disc}

The utility of gyrochronology-based age inferences for hot Jupiter
host stars has been questioned both theoretically and empirically
\citep[e.g.,][]{barnes2007,brown2014,tajeda2021}.  Similarily, we argue
that gyrochronology-based age inferences for WASP-77 A are unreliable
because of the possibility of tidally mediated angular momentum exchange
between the system's orbital angular momentum $L_{\text{orb}}$ and the
rotational angular momentum of the host star WASP-77 A $L_{\text{rot}}$
(the rotational angular momentum of WASP-77 A b is negligible).  The total
angular momentum $L_{\text{tot}}$ of the WASP-77 A system in its current
configuration is
\begin{eqnarray}
L_{\text{tot},2} & = & L_{\text{orb},2} + L_{\text{rot},2}, \\
L_{\text{orb},2} & = & \left(\frac{M_{\ast} M_{\text{p}}}{M_{\ast} + M_{\text{p}}} \right) \left(\frac{2 \pi}{P_{\text{orb},2}}\right) a_{2}^2 \sqrt{1-e^2}, \\
L_{\text{rot},2} & = & I_{\ast} P_{\text{rot},2}.
\end{eqnarray}
Given the currently observed stellar rotation period $P_{\text{rot},2}
= 15.6 \pm 3.5$ days, the system's orbital period $P_{\text{orb},2} =
1.3600306 \pm 0.0000012$ days, and the system's semimajor axis $a_{2}
= 0.0234 \pm 0.0002$ AU, if we assume a constant stellar momentum of
inertia $I_{\ast}$ with the solar value $I/(M_{\odot} R_{\odot}^2) =
0.070$, the stellar mass and radius we inferred $M_{\ast} = 0.92 \pm
0.02~M_{\odot}$ and $R_{\ast} = 0.96^{+0.01}_{-0.02}~R_{\odot}$, and the
planet mass we inferred $M_{\text{p}} = 1.66 \pm 0.03~M_{\text{Jup}}$,
we find $L_{\text{orb},2} = 2.05 \times 10^{49}$ g cm$^{2}$ s$^{-1}$,
$L_{\text{rot},2} = 3.2 \times 10^{48}$ g cm$^{2}$ s$^{-1}$, and
$L_{\text{tot},2} = 2.4 \times 10^{49}$ g cm$^{2}$ s$^{-1}$.

For our isochrone-inferred age of the system $\tau_{\text{iso}}$
= $7.99^{+1.90}_{-1.71}$ Gyr, the rotational evolution models of
\citet{vansaders2016} predict $P_{\text{rot}} \approx 30$ days for a
system with the zero-age main sequence $T_{\text{eff}}$ expected for
WASP-77 A.  Assuming the conservation of angular momentum over the
system's evolution and taking $P_{\text{rot},1} \approx 30$ days as
the rotation period WASP-77 A would have had in the absence of tidal
evolution, we find that the current rapid rotation of WASP-77 A given
the system's age can be explained if the system had $P_{\text{orb},1}
\approx 1.7$ days and $a_{1} \approx 0.027$ AU before tidal evolution.
That is, the disagreement between the isochrone- and gyrochronology-based
ages for WASP-77 A can easily be explained by tidal evolution mediated
angular momentum exchange.

Making the assumptions that the orbit of the WASP-77 A system had
circularized early in its evolution and that the spin-up of WASP-77
A results from tidally mediated angular momentum exchange after orbit
circularization, one can analytically solve the ordinary differential
equation for semimajor axis evolution from \citet{jackson2008} for WASP-77
A's tidal dissipation parameter $Q_{\ast}'$ \citep[i.e., following the
convention of][]{goldreich1966}.  Taking $a_{1} \approx 0.027$ AU, $a_{2}
\approx 0.023$ AU, and $\Delta t \approx 8$ Gyr we find $Q_{\ast}' \sim
10^{8}$.  In other words, even in the limit of inefficient dissipation
tides can easily explain the spin-up of WASP-77 A due to tidal evolution
over the life of the system.  These results provide further evidence that
gyrochronology-based ages should not be relied on for hot Jupiter systems.

\section{Conclusion}\label{conclusion}

We find that the hot Jupiter host star WASP-77 A has super-solar
carbon $[\text{C/H}]_{\text{3D non-LTE}} = 0.10 \pm 0.09$ and oxygen
$[\text{O/H}]_{\text{3D non-LTE}} = 0.23 \pm 0.02$ abundances plus
sub-solar C/O abundance ratios $[\text{C/O}]_{\text{3D non-LTE}} = -0.13
\pm 0.09$ and $\text{C/O}_{\text{3D non-LTE}} = 0.44^{+0.07}_{-0.08}$.
As reported by \citet{line2021}, the atmosphere of the hot Jupiter
WASP-77 A b has $[\text{C/H}] = -0.46^{+0.17}_{-0.16}$, $[\text{O/H}]
= -0.49^{+0.14}_{-0.12}$, and $\text{C/O} = 0.59 \pm 0.08$.  Though
\citet{line2021} assumed that WASP-77 A has solar carbon and oxygen
abundances and therefore asserted that WASP-77 A b has a stellar
C/O abundance ratio, we find that WASP-77 A b has a significantly
super-stellar C/O abundance ratio.  The non-solar abundance ratios
of WASP-77 A qualitatively change the interpretation of WASP-77
A b's atmospheric abundances put forward in \citet{line2021}.
While \citet{line2021} suggested that WASP-77 A b formed its envelope
interior to its parent protoplanetary disk's H$_{2}$O ice line from
carbon-depleted gas with little subsequent planetesimal accretion or
core erosion, we find that its super-stellar C/O abundance ratio implies
formation outside its parent protoplanetary disk's H$_{2}$O ice line.

While the use of exoplanet atmospheric abundances to inform planet
formation is not always straightforward, this qualitative change in the
interpretation of the atmospheric abundances of WASP-77 A b emphasizes
the importance of exoplanet host star elemental abundances inferences.
For the most powerful constraints, these carbon and oxygen abundances
cannot be inferred independently of stellar and exoplanet parameters.
Instead, photospheric and fundamental stellar parameters that are
self-consistent and physically consistent with stellar evolution should
be used both in elemental abundance inferences and to re-derive planet
parameters from observables like Doppler velocities and light curves.
This approach results in self-consistent photospheric and fundamental
stellar parameters, elemental abundances, and exoplanet parameters that
can be used to explore planet formation and evolution.  Accounting
for the possibility of tidally mediated angular momentum exchange
in the WASP-77 A system resolves the tension between its isochrone-
and gyrochronology-based ages.  This result is further evidence that
in general gyrochronology-based ages should not be relied on for hot
Jupiter systems.

\section*{Acknowledgments}
We thank the referee for a prompt and helpful review.  Henrique
Reggiani acknowledges support from a Carnegie Fellowship.  Based on
observations obtained with the Apache Point Observatory 3.5-meter
telescope, which is owned and operated by the Astrophysical Research
Consortium.  The national facility capability for SkyMapper has been
funded through ARC LIEF grant LE130100104 from the Australian Research
Council, awarded to the University of Sydney, the Australian National
University, Swinburne University of Technology, the University of
Queensland, the University of Western Australia, the University of
Melbourne, Curtin University of Technology, Monash University and the
Australian Astronomical Observatory.  SkyMapper is owned and operated
by The Australian National University's Research School of Astronomy
and Astrophysics.  The survey data were processed and provided by
the SkyMapper Team at ANU.  The SkyMapper node of the All-Sky Virtual
Observatory (ASVO) is hosted at the National Computational Infrastructure
(NCI). Development and support of the SkyMapper node of the ASVO has been
funded in part by Astronomy Australia Limited (AAL) and the Australian
Government through the Commonwealth's Education Investment Fund (EIF)
and National Collaborative Research Infrastructure Strategy (NCRIS),
particularly the National eResearch Collaboration Tools and Resources
(NeCTAR) and the Australian National Data Service Projects (ANDS).
This work has made use of data from the European Space Agency (ESA)
mission {\it Gaia} (\url{https://www.cosmos.esa.int/gaia}), processed
by the {\it Gaia} Data Processing and Analysis Consortium (DPAC,
\url{https://www.cosmos.esa.int/web/gaia/dpac/consortium}).  Funding for
the DPAC has been provided by national institutions, in particular the
institutions participating in the {\it Gaia} Multilateral Agreement.
This publication makes use of data products from the Two Micron All
Sky Survey, which is a joint project of the University of Massachusetts
and the Infrared Processing and Analysis Center/California Institute of
Technology, funded by the National Aeronautics and Space Administration
and the National Science Foundation.  This research has made use of the
NASA/IPAC Infrared Science Archive, which is funded by the National
Aeronautics and Space Administration and operated by the California
Institute of Technology.  Funding for the Sloan Digital Sky Survey IV has
been provided by the Alfred P. Sloan Foundation, the U.S.  Department
of Energy Office of Science, and the Participating Institutions.
SDSS-IV acknowledges support and resources from the Center for High
Performance Computing  at the University of Utah. The SDSS website is
\url{www.sdss.org}.  SDSS-IV is managed by the Astrophysical Research
Consortium for the Participating Institutions of the SDSS Collaboration
including the Brazilian Participation Group, the Carnegie Institution
for Science, Carnegie Mellon University, Center for Astrophysics |
Harvard \& Smithsonian, the Chilean Participation Group, the French
Participation Group, Instituto de Astrof\'isica de Canarias, The Johns
Hopkins University, Kavli Institute for the Physics and Mathematics of
the Universe (IPMU) / University of Tokyo, the Korean Participation Group,
Lawrence Berkeley National Laboratory, Leibniz Institut f\"ur Astrophysik
Potsdam (AIP),  Max-Planck-Institut f\"ur Astronomie (MPIA Heidelberg),
Max-Planck-Institut f\"ur Astrophysik (MPA Garching), Max-Planck-Institut
f\"ur Extraterrestrische Physik (MPE), National Astronomical Observatories
of China, New Mexico State University, New York University, University of
Notre Dame, Observat\'ario Nacional / MCTI, The Ohio State University,
Pennsylvania State University, Shanghai Astronomical Observatory,
United Kingdom Participation Group, Universidad Nacional Aut\'onoma
de M\'exico, University of Arizona, University of Colorado Boulder,
University of Oxford, University of Portsmouth, University of Utah,
University of Virginia, University of Washington, University of Wisconsin,
Vanderbilt University, and Yale University.  This research has made use of
NASA's Astrophysics Data System Bibliographic Services.  This research
has made use of the SIMBAD database, operated at CDS, Strasbourg,
France \citep{wenger2000}.  This research has made use of the VizieR
catalogue access tool, CDS, Strasbourg, France (DOI: 10.26093/cds/vizier).
The original description of the VizieR service was published in 2000,
A\&AS 143, 23 \citep{ochsenbein2000}.  This research has made use of the
NASA Exoplanet Archive, which is operated by the California Institute
of Technology, under contract with the National Aeronautics and Space
Administration under the Exoplanet Exploration Program.


\vspace{5mm}
\facilities{ARC, CDS, Exoplanet Archive, Gaia, IRSA, Skymapper, Sloan,
TESS}

\software{\texttt{astropy} \citep{astropy2013,astropy2018},
          \texttt{CERES} \citep{brahm2017},
	  \texttt{colte} \citep{casagrande2021},
	  \texttt{eleanor} \citep{feinstein2019},
          \texttt{IRAF} \citep{tody1986,tody1993},
          \texttt{isochrones} \citep{morton2015},
	  \texttt{iSpec} \citep{blanco-cuaresma2014,blanco-cuaresma2019},
	  \texttt{lightkurve} \citep{lightkurve2018},
	  \texttt{MOOG} \citep{sneden1973},
          \texttt{MultiNest} \citep{feroz2008,feroz2009,feroz2019},
	  \texttt{numpy} \citep{harris2020},
          \texttt{pandas} \citep{mckinney2010,reback2020},
	  \texttt{PyMultinest} \citep{buchner2014},
	  \texttt{q$^{2}$} \citep{ramirez2014},
	  \texttt{RadVel} \citep{fulton2017,fulton2018},
          \texttt{scipy} \citep{virtanen2020}
          }

\bibliography{ms}{}
\bibliographystyle{aasjournal}

\end{document}